\documentclass[twocolumn,showpacs,preprintnumbers,amsmath,amssymb]{revtex4}
\usepackage{graphicx}
\usepackage{bm}
\def\Vec#1{\mbox{\boldmath $#1$}}
\def\dis{\displaystyle}
\def\itmb{\begin{itemize}}
\def\itme{\end{itemize}}
\def\enmb{\begin{enumerate}}
\def\enme{\end{enumerate}}
\def\eqnb{\begin{equation}}
\def\eqne{\end{equation}}
\def\PTP{Prog. Theor. Phys.(Kyoto)}

\def\NPB{{Nucl. Phys.} {\bf B}}
\def\PLB{{Phys. Lett.} B}

\def\PRD{{Phys. Rev.} D}

\begin{document}
\newcommand{\ttbs}{\char'134}

\title{Infrared features of unquenched finite temperature lattice Landau gauge QCD }
\author{Sadataka Furui}
\email{furui@umb.teikyo-u.ac.jp}
\homepage{http://albert.umb.teikyo-u.ac.jp/furui_lab/furuipbs.htm}
\affiliation{%
School of Science and Engineering, Teikyo University, 320-8551 Japan.
}%
\author{Hideo Nakajima}
\email{nakajima@is.utsunomiya-u.ac.jp}
 
\affiliation{
Department of Information Science, Utsunomiya University, 321-8585 Japan. 
}%

\date{\today}
\begin{abstract}
The color diagonal and color antisymmetric ghost propagators slightly above $T_c$ of $N_f=2$ MILC $24^3\times 12$ lattices are measured and compared with zero temperature unquenched $N_f=2+1$ MILC$_c$ $20^3\times 64$ and MILC$_f$ $28^3\times 96$ lattices and zero temperature quenched $56^4$ $\beta=6.4$ and 6.45 lattices. The expectation value of the color antisymmetric ghost propagator $\phi^c(q)$ is zero but its Binder cumulant, which is consistent with that of $N_c^2-1$ dimensional Gaussian distribution below $T_c$, decreases above $T_c$. Although the color diagonal ghost propagator is temperature independent, the $l^1$ norm of the color antisymmetric ghost propagator is temperature dependent. The expectation value of the ghost condensate observed at zero temperature unquenched configuration is consistent with 0 in $T>T_c$. 

We also measure transverse, magnetic and electric gluon propagator and extract gluon screening masses.  The running coupling measured from the product of the gluon dressing function and the ghost dressing function are almost temperature independent but the effect of $A^2$ condensate observed at zero temperature is consistent with 0 in $T>T_c$.

The transverse gluon dressing function at low temperature has a peak in the infrared at low temperature but it becomes flatter at high temperature.  The magnetic gluon propagator at high momentum depends on the temperature. 
These data imply that the magnetic gluon propagator and the color antisymmetric ghost propagator are affected by the presence of dynamical quarks and there are strong non-perturbative effects through the temperature-dependent color anti-symmetric ghost propagator.
\end{abstract}

                          
\maketitle
\section{Introduction}
As a condition of the color confinement in the infrared region, Kugo-Ojima criterion\cite{KO} and the Gribov-Zwanziger scenario\cite{Gr,Zw} are well-known. In these theories, infrared divergence of the ghost propagator is the essential ingredient of the emergence of the string-like inter-quark potential.
In finite temperature SU(2) lattice Coulomb gauge simulation, linearly rising color-Coulomb potential was observed to remain above the deconfinement temperature $T_c$\cite{GOZ04}. The color-Coulomb potential is defined by the ghost propagator and the color diagonal ghost propagator is found to be essentially temperature independent.  The temperature dependence of the color antisymmetric ghost porpagator was not explored.
Coulomb gauge is a non-covariant gauge and the momentum in the time direction is not affected by the gauge fixing. In finite temperature, the momentum in the time direction is interpreted as the Matsubara frequency with appropriate boundary conditions. 

The Landau gauge is a covariant gauge, and in the analysis of the quenched and unquenched lattice Landau gauge simulation\cite{FN03,FN04}, we analyzed zero temperature $N_f=2+1$ unquenched configuration of the SU(3) MILC collaboration of 1) $20^3\times 64$ $\beta_{imp}=6.76,6.83$, which is denoted as MILC$_c$, since realtively coarse lattice of spacing $a=0.12$fm is used, and 2) $28^3\times 96$ $\beta_{imp}=7.09,7.11$, which is denoted as MILC$_f$, since relatively fine lattice of spacing $a=0.09$fm is used\cite{MILC}.

We observed that the Kugo-Ojima parameters of  MILC$_c$ and MILC$_f$ are consistent with 1 while the quenched configuration of $56^4$ lattice remained about 0.8\cite{FN05b,FN06b}.  The Binder cumulant of the color anti-symmetric ghost propagator of the zero temperature quenched SU(2) and  unquenched $N_f$=2+1, SU(3) configurations of the MILC collaboration were consistent with those the $N_c^2-1$ dimensional Gaussian distribution, where $N_c$ is the number of colors. The dynamical quark has the effect of quenching randomness of the system\cite{FN06a}.

In this paper, we extend the analysis of the color anti-symmetric ghost propagator to the quenched SU(3) $56^4$ lattices of $\beta=6.4$\cite{FN04} and 6.45\cite{FN06b} and the finite temperature $N_f=2$ unquenched SU(3) $24^3\times 12$ configurations of $\beta=5.65, 5.725, 5.85$, produced by 
the MILC collaboration\cite{MILCft}.  We denote the finite temperature configurations as MILC$_{ft}$.  The ghost propagator contributes in the off-shell quark gluon vertex via Ward-Slavnov-Taylor identity\cite{DOS00}, and so we expect the screening masses of the gluons produced by the quark loops and ghost loops would be affected by the ghost propagator. We study the magnetic and the electric screening mass of the gluon of the MILC finite temperature configurations.

In the analysis of \cite{MILCft}, the three configurations correspond to the temperature $T=143, 172.5$ and 185 MeV$_\rho$, respectively where subscript $_\rho$ means that the scale is fixed from the mass of the $\rho$ meson $m_\rho=770$MeV. The temperature $T_c$ at which the chiral susceptibility shows a peak, indicating the cross over to the deconfinement was assigned to be about 140MeV.  Since the standard Wilson plaquette action was used in the production of the gauge configuration, the flavor symmetry was broken and the ratio of the $\rho$ mass to pion mass is larger than the physical value\cite{BKTG95}. Recent simulation with Asqtad action\cite{MILC06} suggests that $T_c\sim 170$MeV, consistent with the value of that of $N_f=2$ improved Kogut-Susskind (KS) fermion $T_c\sim 173\pm 8$MeV \cite{KaLa03}. A systematic comparison of the $\rho$ meson mass for improved staggered actions in quenched approximation is given in \cite{ccjkmpp06}.
Recently, \cite{AFKS06} claims that the transition temperature depends on which physical quantity one measures, and that the $T_c$ defined by the chiral susceptibility is 151(3)MeV and consistent within errors with \cite{MILCft}. They showed that the $T_c$ defined by the strange quark chiral susceptibility and the Polyakov loop are about 175MeV and consistent with \cite{KaLa03}.

 Since the continuum limit of the mass of the vector meson would not depend on the temperature near $T_c$\cite{Karsch95}, it would be natural to assign the bare lattice $\rho$ mass by about 20\% heavier and shift the temperature by the same amount. We leave a more accurate assignment of the temperature scale of the MILC$_{ft}$ to a future study and since the disconnected chiral susceptibility measured by the configurations shows a clear peak at around 140MeV$_\rho$, we assign the $\beta=5.65, 5.725$ and 5.85 data, which correspond to $T=143, 172.5$ and 185 MeV$_\rho$, by $T/T_c=1.02, 1.23$ and 1.32, respectively.
We compare quenched and unquenched ghost propagator of zero temperature and finite temperature and investigate the role of quarks on the gluon field and its dependence on the temperature.

The organization of the paper is as follows. In the Sect.2, we show the results of the color diagonal ghost propagator of the quenched $56^4$ and unquenched finite temperature MILC configurations. The corresponding color antisymmetric ghost propagators are shown in the Sect.3. The Kugo-Ojima parameters, the transverse, magnetic and electric gluon propagator and the QCD running coupling of the finite temperature unquenched configurations are shown in the Sects.4, 5 and 6, respectively.  Conclusions and a discussion are given in the Sect.7.

\section{The color diagonal ghost propagator}

The ghost propagator is defined by the Fourier transform(FT) of the  expectation value of the inverse Faddeev-Popov(FP) operator ${\mathcal  M}=-\partial_\mu D_\mu$, where $D_\mu$ is the covariant derivative, as
\begin{eqnarray}
FT[D_G^{ab}(x,y)]&=&FT\langle {\rm tr} ( \Lambda^{a \dagger} \{({\cal  M}[U])^{-1}\}_{xy}
\Lambda^b)  \rangle,\nonumber\\
&=&\delta^{ab}D_G(q^2).
\end{eqnarray}
where $U$ is the link variable obtained by the Landau gauge fixing. In all the simulation in this work we adopt the $\log U$ definition of the gauge field, 
and the SU(3) color matrix $\Lambda$ is normalized as ${\rm tr}\Lambda^a\Lambda^b=\delta^{ab}$. Number of samples is about 10 each in the $56^4$ lattices and about 100 each in the $24^3\times 12$ lattices.

In the usual Monte Carlo simulation, it is necessary to make an average over many samples, but the color diagonal ghost propagator and the color anti-symmetric ghost propagator of unquenched configurations are almost independent of samples. Since we adopt the cylinder cut, i.e. select the momentum along the diagonal direction in momentum space, and take into account translation invariance and rotational symmetry of the lattice, the error bar of the ghost propagator of one sample is already not so large, and an order of 10 samples is enough to get the expectation values, although one should be cautious to the appearance of exceptional samples\cite{FN04}. 

The color diagonal ghost propagator is defined as
\begin{eqnarray}
&&D_G(q)=\frac{1}{N_c^2-1}\frac{1}{V}\nonumber\\
&&\times {\rm tr}\left\langle \delta^{ab}(\langle \Lambda^a\cos{ q x}|f_c^b({ x})\rangle+\langle \Lambda^a\sin{ q x}|f_s^b({ x})\rangle)\right\rangle\nonumber\\
&&=G(q^2)/q^2\nonumber
\end{eqnarray}
Here ${f_c}^b({ x})$ and ${f_s}^b({ x})$ are the solution of ${\mathcal M} f^b({ x})=\rho^b({ x})$ with $\displaystyle \rho^b({ x})=\frac{1}{\sqrt V}\Lambda^b\cos{q x}$ and $\displaystyle \frac{1}{\sqrt V}\Lambda^b\sin{ q x}$, respectively.  

\subsection{Quenched SU(3) $56^4$ lattice}

\begin{figure}[htb]
\begin{center}
\includegraphics[width=7cm,angle=0,clip]{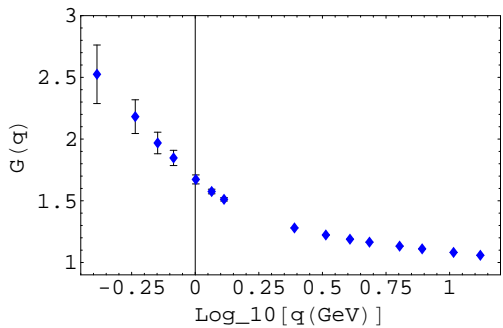}
\end{center}
\caption{The ghost dressing function of quenched $56^4$ configurations of  $\beta=6.45$ as the function of $\log_{10}$[q(GeV)]. }\label{645_ghd}
\end{figure}

The color diagonal ghost propagators of $56^4$ quenched SU(3) are shown in \cite{FN04}, but we show the data for a comparison with the color anti-symmetric ghost propagator in the next subsection. The FIG. \ref{645_ghd} is the ghost dressing function, of the $56^4$ $\beta=6.45$ configurations produced by the Monte Carlo simulation and subsequently Landau gauge fixed.

The scale of the $\beta=6.4$ and 6.45 configurations are fixed as in TABLE \ref{scale}, using the formula in \cite{NeSo01}.
The ghost propagator is infrared divergent, but its singularity is weaker than $q^{-4}$.

\begin{table}
\begin{center}
\caption {The $\beta=2N_c/{g_0}^2$  ($\beta_{imp}=5/3\times \beta$ for MILC$_f$ and MILC$_c$), the inverse lattice spacing $1/a$, lattice size and lattice length(fm) of configurations investigated in this paper. Subscripts $c$ and $f$ of MILC correspond to coarse lattice($a$=0.12fm) and fine lattice($a$=0.09fm).  The MILC$_{ft}$ corresponds to the $N_f=2$ finite temperature configurations and we assign $\beta=5.65, 5.725$ and 5.85 data to $T/T_c=1.02, 1.23$ and 1.32, respectively.}\label{scale}
\begin{tabular}{c c c c c c c c}
   &$\beta_{imp}/\beta$ &  $am_{ud}/am_{s}$& $N_f$& $1/a$(GeV)&$L_s$ & $L_t$ &$a L_s$(fm)\\
\hline  
quench & 6.4 &             & 0 & 3.66   & 56 & 56  & 2.96\\
       & 6.45 &            & 0 & 3.87 & 56 & 56  & 2.94\\
\hline
MILC$_c$ &6.83&       0.040/0.050& 2+1 & 1.64 & 20 &64&2.41\\
       &6.76&       0.007/0.050& 2+1 & 1.64 & 20 &64&2.41\\
\hline
MILC$_f$ &7.11&       0.0124/0.031& 2+1 & 2.19 &28 & 96&2.52\\
       &7.09 &       0.0062/0.031& 2+1 & 2.19 &28 & 96&2.52\\
\hline
MILC$_{ft}$ &5.65&       0.008& 2 & 1.716 &24 & 12&  2.76   \\
          &5.725&       0.008& 2 & 1.914 &24 & 12&  2.47   \\
          &5.85&       0.008& 2 & 2.244 &24 & 12&  2.11   \\
\hline
\end{tabular}
\end{center}
\end{table}

\subsection{MILC finite temperature $24^3\times 12$ lattice}

\begin{figure}[htb]
\begin{center}
\includegraphics[width=7cm,angle=0,clip]{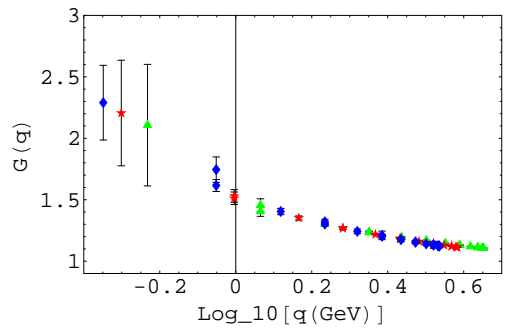}
\end{center}
\caption{The ghost dressing function of MILC$_{ft}$ configurations of  $T/T_c=1.02$(blue diamonds),  $T/T_c=1.23$(red stars) and $T/T_c=1.32$(green triangles)
as the function of $\log_{10}$[q(GeV)].  }\label{milcft_ghd}
\end{figure}

The color diagonal ghost dressing function of MILC finite temperature configurations are shown in FIG. \ref{milcft_ghd}.  We observe that the three data of different temperatures scale as shown in the figure when each lattice spacing $a$ is chosen as in the ref.\cite{MILCft}. 

\section{The color antisymmetric ghost propagator}

The color anti-symmetric ghost propagator is defined as
\begin{eqnarray}
&&\phi^c(q)=\frac{1}{\mathcal N}\frac{1}{V}\nonumber\\
&&\times {\rm tr}\left\langle f^{abc}(\langle \Lambda^a\cos{q x}|f_s^b({ x})\rangle-\langle \Lambda^a\sin{ q x}|f_c^b({ x})\rangle)\right\rangle\nonumber
\end{eqnarray}
where the outer-most bracket means the ensemble average.

In a theory based on the Curci-Ferrari gauge and in its extension to the Landau gauge, a parameterization of the color anti-symmetric ghost propagator with use of the ghost condensate parameter $v$ was proposed\cite{Gent03}.

A simulation in the SU(2) lattice Landau gauge using a parameterization
\begin{equation}\label{phifit}
 \frac{1}{N_c^2-1}\sum_a|\phi^a(q)|= \frac{r/L^2+v}{q^4+v^2}
\end{equation}
is performed in \cite{CMM05}. In \cite{Gent03} the ghost propagator in the condensed vacuum in SU(2) is expressed as
\begin{equation}
\langle \bar c^a c^b\rangle_q=-i\frac{q^2\delta^{ab}-v\epsilon^{ab}}{q^4+v^2}.
\end{equation}
The term $r/L^2$ in eq.(\ref{phifit}) is a term introduced to incorporate the finite lattice size effect.  We applied the method to zero temperature unquenched SU(3) configurations of MILC$_c$ and MILC$_f$\cite{FN06a,FN06b}.  
In \cite{CMM05}, finite size effects expressed by $r/L^2$ is disentangled from the fitting of the lattice data of color antisymmetric ghost propagator in high momentum region, where the perturbative QCD (pQCD) is a good approximation, as
\begin{equation}\label{ghostscale}
\frac{1}{N_c^2-1}\sum_a\frac{L^2}{\cos(\pi \bar q/L)}|\phi^a(q)|=\frac{r}{q^z},
\end{equation}
where the denominator in $\displaystyle \frac{|\phi^a(q)|}{\cos(\pi\bar q/L)}$ is the factor that appears in the vertex function of the lattice perturbation theory and $\bar q$ takes integer values $0,1,\cdots L/2$. In the present work, 
since the high momentum data points where the enhancement due to $1/\cos$ factor is significant are few, i.e. the enhancement of the highest momentum point of FIG.\ref{phi_709} is 0.64, that of the next to highest momentum point is 0.35 and the rest are less than 0.2 in the $\log_{10}$ scale, we fix the parameter $r$ and $v$ by the fit of $|\phi^a(q)|$ using eq.(\ref{phifit}) including the whole momentum region except the highest momentum point.

This treatment of the finite size effect on the lattice data is not without ambiguity, but our aim is to search qualitative differences as the temperature or the masses of quarks are changed.

Another quantity that characterizes the system is the Binder cumulant of the color antisymmetric ghost propagator defined as
\[
U(q)=1-\frac{\langle\vec\phi(q)^4\rangle}{3\langle\vec\phi(q)^2\rangle^2}.
\]

A simulation of SU(2) lattice Landau gauge obtained $U\sim 0.44$, almost independent of the momentum. This value is compatible with that of the three dimensional Gaussian distribution\cite{FN06a} and the analysis of SU(3) MILC$_c$ and MILC$_f$ showed that $U$ is compatible with that of eight dimensional Gaussian distribution.  
We extend these analyses to large quenched lattices and the finite temperature configurations.

\subsection{Quenched SU(3) $56^4$ lattice}

\begin{figure}
\begin{center}
\includegraphics[width=7cm,angle=0,clip]{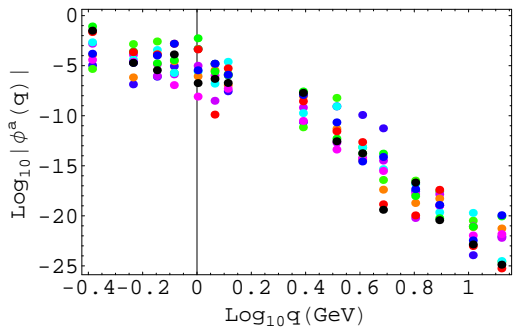}
\end{center}
\caption{$\log_{10}|\vec\phi(q)|$ as the function of  $\log_{10}q$(GeV) of quenched $\beta=6.45, 56^4$ lattice. }\label{gh645_anti}
\end{figure}

The absolute value of the color antisymmetric ghost propagator of quenched $56^4$ lattice in the infrared region is about 3 orders of magnitude smaller than that of the color diagonal ghost propagator and the values are sample dependent. Results of $\beta=6.45, 56^4$ 10 samples are shown in FIG.\ref{gh645_anti}.

Due to this sample dependence, quite different from the case of unquenched  $20^3\times 64$ lattices\cite{FN06a} and $28^3\times 96$\cite{FN06b}, the Binder cumulant of the color antisymmetric ghost propagator of quenched configurations is noisy due to large  $\langle\vec\phi(q)^4\rangle$ as compared to $\langle\vec\phi(q)^2\rangle^2$. The randomness of the color anti-symmetric ghost propagator of quenched configurations is large and the Binder cumulant becomes unstable.

\subsection{MILC zero temperature $28^3\times 96$ lattice}

\begin{figure}
\begin{center}
\includegraphics[width=7cm,angle=0,clip]{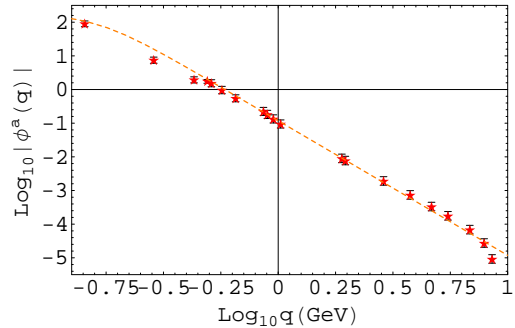}
\end{center}
\caption{$\log_{10}|\vec\phi(q)|$ as the function of  $\log_{10}q$(GeV) of MILC$_f$ and the fit using $r=134$ and $v=0.026$GeV$^2$. }\label{phi_709}
\end{figure}

The color antisymmetric ghost propagator of MILC$_f$ is shown in FIG. \ref{phi_709}. The fitting parameters are given in \cite{FN06b}. 

The expectation value of $v$ is small but the condition that the fitted curve passes the lowest momentum point of FIG.\ref{phi_709} within the error bar gives $v=0.026(6)$GeV$^2$. It suggests that the presence of BRST partner of $A^2$ condensate at zero temperature. 

The Binder cumulant of the zero temperature $N_f=2+1$ MILC$_c$ lattice is reported in \cite{FN06a} and that of MILC$_f$ is reported in \cite{FN06b}. We observed that the mass function of $m_0=27.2$MeV quark propagator of $\beta_{imp}=7.09$ with bare mass combination $m_0=27.2$MeV/68MeV shows an anomalous behavior in $q<1$GeV region, and that in the same region Binder cumulant $U(q)$ shows an anomalous behavior, although the mass function of $m_0=68$MeV does not show the anomaly.  The non-QCD like behavior of staggered quarks calculated with large lattice spacing $a$  and small bare mass $m_0$ is reported in \cite{HH06}.  Since no anomaly was observed in $\beta=7.09$ $m_0=13.6$MeV/68MeV \cite{FN06a,FN06b}, effects of the relative size of the $s-$quark mass v. $ud-$quark mass and the number of $N_f$ are suggested.

Thus we extend the analysis to MILC configuration of $N_f=3$, $\beta_{imp}=7.18$, $28^3\times 96$ lattice with the bare quark mass $m_0=0.031$ and that of $N_f=2$, $\beta_{imp}=7.20$, $20^3\times 64$ lattice with the bare quark mass $m_0=0.02$\cite{MILC_2f3f}.   A result of the Binder cumulants of the $N_f=3$ and the $N_f=2$ (50 samples) are 0.66(1), i.e. $\langle \vec \phi(q)^4\rangle$ is close to $\langle \vec \phi(q)^2\rangle^2$. When the bare masses of the quarks are the same, the system possesses the self averaging property\cite{LaBi05}, and when they are different as in $N_f=2+1$, the system lacks this property.  The parameter $v$ fitted from $N_f=2$ (50 samples) is found to be consistent with 0.

\subsection{MILC finite temperature $24^3\times 12$ lattice}

\begin{figure}[htb]
\begin{center}
\includegraphics[width=7cm,angle=0,clip]{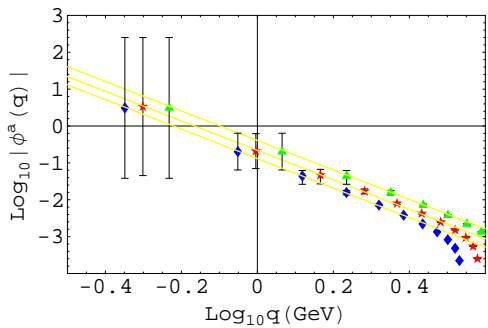}
\end{center}
\caption{$\log_{10}|\vec\phi(q)|$ as the function of $\log_{10}\Vec q(GeV)$ of MILC$_{ft}$ of  $T/T_c=1.02$(blue diamonds),  $T/T_c=1.23$(red stars) and $T/T_c=1.32$(green triangles). }\label{milcft_ghlanti}
\end{figure}

The fluctuation of the color antisymmetric ghost propagator around the expectation value 0 was almost Gaussian in the case of zero temperature unquenched configurations\cite{FN06a,FN06b}.
The logarithm of the absolute value of the color antisymmetric ghost propagator of finite temperature unquenched configurations as a function of the logarithm of the momentum is shown in the FIG. \ref{milcft_ghlanti}.  In contrast to the color diagonal ghost propagator, the absolute value of the color antisymmetric ghost propagator depends on the temperature. The temperature dependence of the scale of $\phi$ defined by $r$ of eq.(\ref{ghostscale}) can be expressed as roughly $r\sim 5.49a^{-4.23}$, where $a$ is the lattice spacing at each temperature in the unit of GeV.

The fitting parameters of the color anti-symmetric ghost propagator are given in TABLE \ref{ghstcdst}.  The condensate parameter $v$ of finite temperature is difficult to assign since the lattice spacing is relatively large and difficult to detect the infrared bending behavior in the color antisymmetric ghost propagator, but they are consistent with 0. 

\begin{table}[htb]
\caption{The fitted parameters $r,z$ and $v$ of the color antisymmetric ghost propagator $|\phi(q)|$ of MILC$_c$, MILC$_f$ and MILC finite temperature. Two values of $U$ of MILC$_f$ correspond to the average below $q=1$GeV and the average above 1GeV, respectively.}\label{ghstcdst}
\begin{center}
\begin{tabular}{c c c c c   c c c   c}
$\beta_{imp}/\beta$ & $m_0$(MeV)& $r$   &$z$ &$v$ &  $U$\\
\hline
6.76 & 11.5/82.2  &  37.5 & 3.90& 0.02(1) &  0.53(5)\\
6.83 & 65.7/82.2  &  38.7 & 3.85& 0.01(1) &  0.57(4)\\
\hline
7.09 & 13.6/68.0  & 134 & 3.83 & 0.026(6) & 0.57(4)/0.56(1)\\
7.11 & 27.2/68.0  & 112 & 3.81 & 0.028(8) & 0.58(2)/0.52(1)\\
\hline
5.65 &     12.3    &  54.4   &  4.01    &  0.0    & 0.580(13)     \\
5.725&     12.8    &  88.3   &  3.95    &  0.0    & 0.571(4)     \\
5.85 &     15.0    &  165.9  &  3.93    &  0.0    & 0.558(2)     \\
\hline
\end{tabular}
\end{center}
\end{table}

\begin{figure}[htb]
\begin{center}
\includegraphics[width=7cm,angle=0,clip]{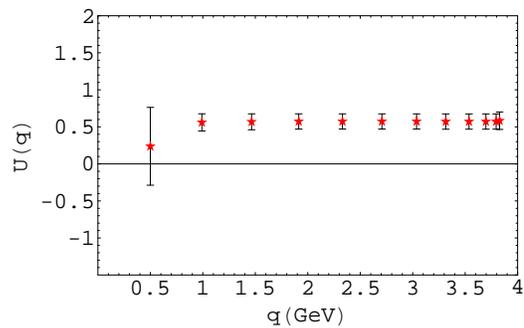}
\end{center}
\caption{The Binder cumulant of the color antisymmetric ghost propagator of MILC $N_f=2$ configurations of $T/T_c=1.23$(red stars). }\label{bindft_03}
\end{figure}

The Binder cumulant of $\phi$ is almost independent of the momentum except the lowest momentum point. We show a typical example of $T/T_c=1.23$ in FIG. \ref{bindft_03}.  

\begin{figure}[htb]
\begin{center}
\includegraphics[width=7cm,angle=0,clip]{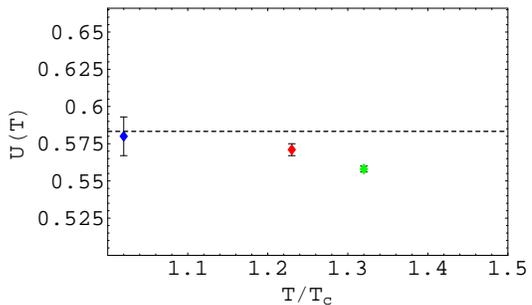}
\end{center}
\caption{Averages over momenta excluding the lowest momentum point of the Binder cumulants of MILC finite temperature configurations. $T/T_c=1.02$(blue diamonds),  $T/T_c=1.23$(red stars) and $T/T_c=1.32$(green triangles). }\label{milcft_bind}
\end{figure}

The temperature dependence of the average of $U(q)$ excluding the lowest momentum point is shown in  FIG. \ref{milcft_bind}. The average $U(T)$ decreases monotonically as a function of $T$ from that of the Gaussian distribution (dashed line) at high temperature.

\section{The Kugo-Ojima color confinement parameter}

The Kugo-Ojima parameter is defined by the two point function of the
covariant derivative of the ghost and the commutator of the antighost and gauge field
\begin{eqnarray}
&&\left(\delta_{\mu\nu}-\frac{q_\mu q_\nu}{q^2}\right)u^{ab}(q^2)\nonumber\\
&&=\frac{1}{V}
\sum_{x,y} e^{-iq(x-y)}\left\langle {\rm tr}\left({\Lambda^a}^{\dag}
D_\mu \displaystyle{\frac{1}{-\partial D}}[A_\nu,\Lambda^b] \right)_{xy}\right\rangle.\nonumber\\ \label{kugou}
\end{eqnarray}
Kugo and Ojima\cite{KO} showed that $u(0)=-1$ is a condition of the color confinement. Zwanziger\cite{Zw} defined the horizon function $h$ that is related to the Kugo-Ojima parameter $c=-u(0)$ as follows.
\[
\sum_{x,y} e^{-iq(x-y)} \left \langle {\rm tr}\left({\Lambda^a}^{\dag}
D_\mu \displaystyle{\frac{1}{-\partial D}}(-D_\nu)\Lambda^b\right)_{xy}\right
\rangle
\]
\[
=G_{\mu\nu}(q)\delta^{ab}
=\left(\displaystyle{\frac{e}{d}}\right)\displaystyle{\frac{q_\mu q_\nu}{q^2}}\delta^{ab}
-\left(\delta_{\mu\nu}-\displaystyle{\frac{q_\mu q_\nu}{q^2}}
\right)u^{ab},
\]
where, with use of the covariant derivative $D_{\mu}(U)$
\[
D_{\mu}(U_{x,\mu})\phi=S(U_{x,\mu})\partial_\mu \phi+[A_{x,\mu},\bar \phi],
\]
$
\partial_\mu \phi=\phi(x+\mu)-\phi(x)$,  
$\bar \phi=\dis\frac{\phi(x+\mu)+\phi(x)}{2}$
and $\displaystyle S(U_{x\mu})=\frac{adj A_{x,\mu}/2}{{\rm tanh} (adj A_{x,\mu}/2)}$.

Using the definition 
\[
e=\left\langle\sum_{x,\mu}{\rm tr}(\Lambda^{a\dag} 
S(U_{x,\mu})\Lambda^a)\right\rangle/\{(N_c^2-1)V\},
\]
the horizon condition reads $\displaystyle \lim_{q\to 0}G_{\mu\mu}(q)-e=0$,
and the left hand side of the condition is 
$
\left(\displaystyle{\frac{e}{d}}\right)+(d-1)c-e=(d-1)h
$
where $h=c-\dis{\frac{e}{d}}$ and dimension $d=4$, and it follows that 
$h=0 \to$ horizon condition, and thus the horizon condition coincides
with Kugo-Ojima criterion provided the covariant derivative approaches
the naive continuum limit, i.e., $e/d=1$.

The Kugo-Ojima parameter is defined by a scalar function at vanishing momentum in the continuum theory, but in the lattice simulation, we measured the magnitude of  the right hand side of the eq.(\ref{kugou}) with $\mu=\nu$ polarizations of $A_\nu$ and $D_\mu$, and then observed in the case of asymmetric lattices there is a strong positive correlation between the magnitude and the lattice size of the axis whose directions are perpendicular to the polarization.
This general features are manifested in the data of MILC$_f$ and MILC$_c$, whose time (4th)
axis is longer than the spatial axes and in the data of MILC$_{ft}$ whose time  axis is shorter than the spatial axes.

\subsection{Quenched SU(3) $56^4$ lattice}
\begin{figure}[htb]
\begin{center}
\includegraphics[width=7cm,angle=0,clip]{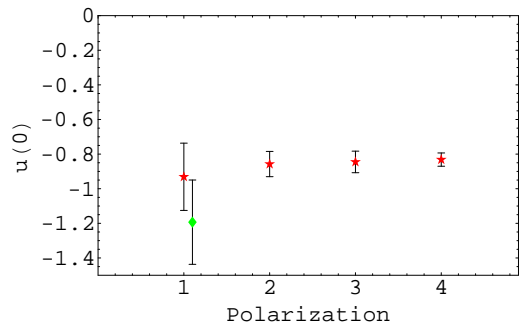}
\end{center}
\caption{Kugo-Ojima parameter $u(0)$ of quenched $56^4$ configurations of $\beta=6.45$(red stars). The data of an exceptional sample is indicated by the green diamond. Polarizations 1,2,3,4 correspond to $x,y,z$ and $t$.}\label{kg_645e}
\end{figure}

The Kugo-Ojima confinement parameter of quenched configuration saturates at about 80\% of the expected value $c=1$. There appear exceptional samples with an average consistent with $c=1$ within errors. The polarization dependence of the Kugo-Ojima parameter of $\beta=6.45$ samples shown in FIG. \ref{kg_645e} is due to the lack of rotational invariance of our system.

\subsection{MILC finite temperature $N_f=2$, $24^3\times 12$ lattice}

Shown in TABLE \ref{kugo} are the Kugo-Ojima parameters and the horizon function deviation of quenched $56^4$ configurations, zero-temperature unquenched MILC$_c$ and MILC$_f$ configurations and unquenched finite temperature MILC$_{ft}$ configurations.

The Kugo-Ojima parameters of unquenched zero-temperature configurations are consistent with the theory\cite{FN05b,FN06a,FN06b}, and those of unquenched finite temperature configurations show temperature dependence.

\begin{figure}[htb]
\begin{center}
\includegraphics[width=7cm,angle=0,clip]{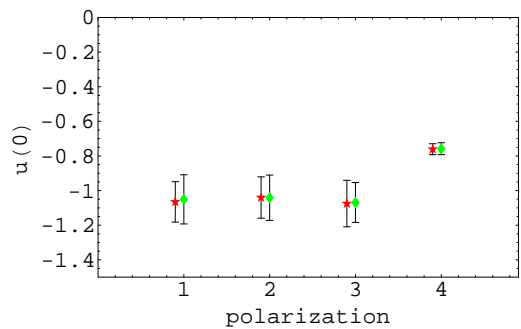}
\end{center}
\caption{Kugo-Ojima parameter $u(0)$ of MILC$_f$ $N_f=2+1$ KS fermion unquenched configurations of $\beta_{imp}=7.11$(green diamonds),  $\beta_{imp}=7.09$(red stars) . }\label{kg_2896}
\end{figure}

\begin{table*}[htb]
\begin{tabular}{c c c c c c c}
\hline
&$\beta_{imp}/\beta$ & $c_x$     & $c_t$    &$c$ &  $e/d$        &    $h$   \\
\hline
quench &6.4  & & &0.827(27)&0.954(1) & -0.12\\
       & 6.45  & &  &0.814(89) &0.954(1) & -0.14\\
\hline
MILC$_c$&6.76 & 1.04(11)  & 0.74(3) &0.97(16) & 0.9325(1) & 0.03(16)  \\
&6.83 & 0.99(14)  & 0.75(3) &0.93(16) & 0.9339(1) &  -0.00(16) \\
\hline
MILC$_f$&7.09 & 1.06(13)  & 0.76(3) &0.99(17) & 0.9409(1) & 0.04(17)  \\
  &7.11 &1.05(13)   &  0.76(3)& 0.98(17) & 0.9412(1) &  0.04(17) \\
\hline
MILC$_{ft}$&5.65 & 0.72(13)  & 1.04(23) &  0.80(21) & 0.9400(7) & -0.14(21) \\
  &5.725  &  0.68(15)& 0.77(16) & 0.70(15) & 0.9430(2) & -0.24(15)\\
  &5.85   &  0.63(19)& 0.60(12) & 0.62(17)  & 0.9465(2) & -0.33(17)\\
\hline
\end{tabular}
\caption{The Kugo-Ojima parameter of the quenched $56^4$ lattice and that of the unquenched KS fermion (MILC$_c$,MILC$_f$,MILC$_{ft}$).  $c_x$ is the polarization along the spatial directions,  $c_t$ is that along the time direction,  $c$ is the weighted average of $c_x$ and $c_t$ i.e. $(3c_x+c_t)/4$, $e/d$ is the trace divided by the dimension and $h$ is the horizon function deviation.}\label{kugo}
\end{table*}

\begin{figure}[htb]
\begin{center}
\includegraphics[width=7cm,angle=0,clip]{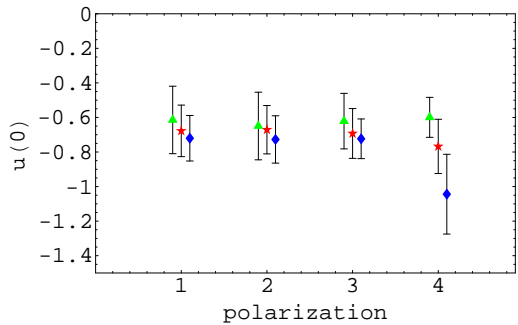}
\end{center}
\caption{Kugo-Ojima parameter $u(0)$ of MILC$_{ft}$ configurations of $T/T_c=1.02$(blue diamonds),  $T/T_c=1.23$(red stars) and $T/T_c=1.32$(green triangles). }\label{milcft_kgplt}
\end{figure}

 The FIG. \ref{kg_2896} show the dependence on the polarization of the Kugo-Ojima parameter. The polarization dependence of the parameter $c$ is a result of an integration over the axes perpendicular to the polarization. It becomes large when there is a long axis perpendicular to the polarization.
The FIG. \ref{milcft_kgplt} shows the temperature and the polarization dependence of the Kugo-Ojima parameter.

 The quenched configurations and high temperature configurations show larger deviation from $u(0)=-1$, which would be due to higher randomness of these systems. The origin of the randomness in the latter would be the thermal fluctuation and that of the former would be lack of fermions that quench randomness of the system.

\section{The finite temperature gluon propagator}

The fluctuation of the gluon propagators of lattice Landau gauge is larger than that of the ghost propagator. We measure the gluon propagators of the three temperatures of MILC$_{ft}$ by using about 100 samples each and choosing the momenta along the three dimensional (3-d) coordinate axes and along the diagonal in the 3-d space. The dependence on the choice of the direction is not significant except a slight suppression of the $\bar q=(1,1,1,0)$, as compared to that with the same magnitude but along the coordinate axes. We fix the normalization of the gluon propagator such that the running coupling defined by the product 
of the gluon dressing function and the ghost dressing function squared as 
defined in Sect.\ref{runningcoupling} in the
high momentum region agrees with the pQCD result and that the rescaling factor 
for the ghost dressing function and the gluon dressing function are the same.
The renormalization factors defined by the highest momentum point along the coordinate axes of the three temperatures are 0.48(1). 

\begin{figure}[htb]
\begin{center}
\includegraphics[width=7cm,angle=0,clip]{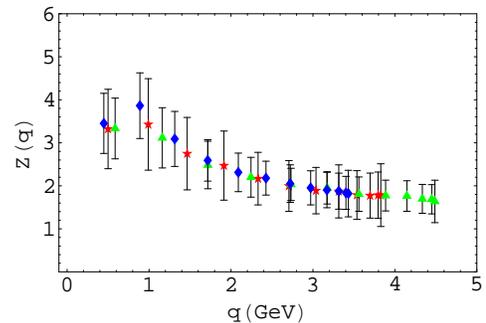}
\end{center}
\caption{The transverse gluon dressing function $Z(q)$ of MILC$_{ft}$ $N_f=2$ KS fermion unquenched configuration of $T/T_c=1.02$(blue diamonds),  $T/T_c=1.23$(red stars) and $T/T_c=1.32$(green triangles). }\label{milcft_Z123}
\end{figure}

In Fig.\ref{milcft_Z123} the transverse gluon dressing functions of the three temperatures taken at momenta along the diagonal direction in the 3-d momentum space after the rescaling are shown. The asymptotic value is about 2, since we normalized Tr$\Lambda^a \Lambda^b=\delta^{ab}$. Larger infrared enhancement of the transverse gluon dressing function for low temperature agrees with
a recent result of quenched SU(2) finite temperature simulation\cite{CMM07}.

The temperature dependence of unquenched transverse gluon propagator and that of the magnetic gluon propagator $G_m(z)$, and the temperature dependence of the quenched SU(2) transverse gluon propagator\cite{CMM07} is similar to that of unquenched SU(3) electric gluon propagator $G_e(z)$, which is shown in Fig.\ref{milcft_gm} and \ref{milcft_ge}, respectively. 

\begin{figure}[htb]
\begin{center}
\includegraphics[width=7cm,angle=0,clip]{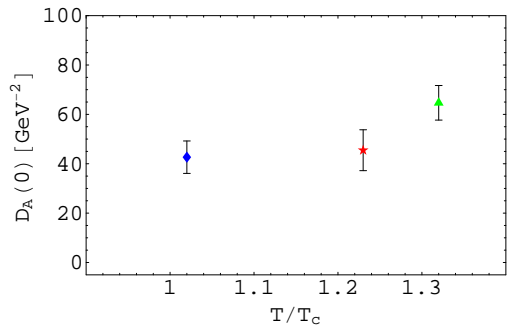}
\end{center}
\caption{The transverse gluon propagator of MILC$_{ft}$ $N_f=2$ KS fermion unquenched configuration of $T/T_c=1.02$(blue diamonds),  $T/T_c=1.23$(red stars) and $T/T_c=1.32$(green triangles) at $q=0$ in the unit of GeV$^{-2}$. }\label{milcft_DA0}
\end{figure}

\begin{figure}[htb]
\begin{center}
\includegraphics[width=7cm,angle=0,clip]{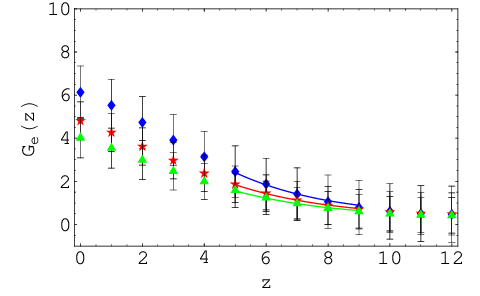}
\end{center}
\caption{Electric gluon propagator of MILC$_{ft}$ $N_f=2$ KS fermion unquenched configuration of $T/T_c=1.02$(blue diamonds),  $T/T_c=1.23$(red stars) and $T/T_c=1.32$(green triangles). }\label{milcft_ge}
\begin{center}
\includegraphics[width=7cm,angle=0,clip]{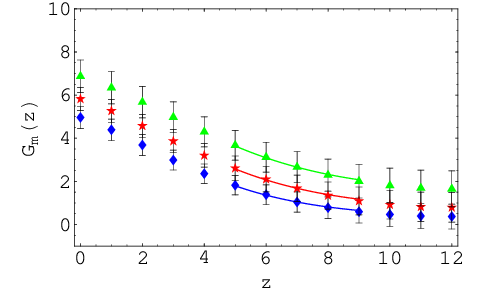}
\end{center}

\caption{Magnetic gluon propagator of MILC$_{ft}$ $N_f=2$ KS fermion unquenched configurations of $T/T_c=1.02$(blue diamonds),  $T/T_c=1.23$(red stars) and $T/T_c=1.32$(green triangles). }\label{milcft_gm}
\end{figure}

The transverse gluon propagator at the zero momentum is finite as shown in Fig.\ref{milcft_DA0}. The value of $D_A(0)$ of MILC$_f$, $\beta=7.09, 28^3\times 96$ is about 15 GeV$^{-2}$\cite{FN05b}. The $D_A(0)$ decreases monotonically as $T$ decreases to 0. Whether it vanishes or remains finite is an important problem for fixing the nature of the infrared fixed point of the running coupling. There is an argument that the infrared non-vanishing of the Landau gauge gluon propagator is a finite size effect. Our lattice volume of low temperature data is larger than the high temperature data, but the running coupling which is calculated by the gluon propagator and shown in Sect. VI excludes large temperature dependence in the finite size effects. A correction of the global scaling factor is, however, not excluded.

We observe that the fluctuation of the gluon propagator at the zero momentum 
increases as the temperature increases.@

The one-loop off-shell contribution to the quark-gluon vertex,
is related to the quark-quark-ghost-ghost amplitude via Ward-Slavnov-Taylor 
identity\cite{DOS00}.
The screening mass of the gluon would be affected by the ghost propagator.
The correlation function of the gauge fields in the position space is defined as \cite{HKR95}
\[
D_{A\mu}(z)=\langle A_\mu(z)A_\mu(0)\rangle,
\]
where $A_\mu(z)=\sum_{x_0,x_1,x_2}A_\mu(x_0,\vec x)$.

The screening mass of the gluon is produced by the quark-, gluon- and ghost-loops. 
The electric screening mass $m_e$ is defined by
\[
G_e(z)=D_{A0}(z)\sim exp[-m_e z/N_t T]
\]
The magnetic screening mass $m_m$ is defined by
\[
G_m(z)=(D_{A1}(z)+D_{A2}(z))/2\sim exp[-m_m z/N_t T]
\]
We fit the propagator $G_e(z)$ and $G_m(z)$ by
\[
\displaystyle D_A(z)=A \cosh[(m/N_t T)(z-N_x/2)].
\]
The mass depends on the region of $z$ used to fit the data and Karsch et al. proposed to fit data of $z$ near $N_x/4$.\cite{HKR98} We fit the data from $z=5$ to 9.  When we fit the data from $z=0$ to $z=6$, the masses become smaller by about 30-40\%. Since deviations from the $\cosh$ functional form become larger in this region, we adopt the fit from $z=5$ to 9.

\begin{table}[htb]
\begin{center}
\caption{The electric and the magnetic screening mass of MILC$_{ft}$.}
\begin{tabular}{c c c c}
$\beta$ &  $T/T_c$ &$m_e/T$ &  $m_m/T$\\ 
\hline
5.65 & 1.02 &  3.42(27) &  3.48(48)  \\
5.725& 1.23 &  3.22(40)  &  2.90(20)  \\
5.85 & 1.32 &  3.14(33)  &  2.31(22)  \\
\hline
\end{tabular}
\end{center}
\end{table}

In high temperature, pQCD suggests that
\[
\frac{m_e}{T}\propto g(T),  \frac{m_m}{T}\propto g^2(T)
\]
where $g^2(T)/4\pi$ is the running coupling at temperature $T$ and 0 momentum.
 Our electric screening mass and the magnetic screening mass are close to those of quenched SU(2) $m_e/T=2.484(52)$ measured by \cite{HKR95}. The magnetic screening mass in the case of SU(2) using the data in the region of $T>2T_c$ was $m_m/T=0.466(25)g^2(T)$ in the two-loop pQCD calculation with $\Lambda_m=0.262(18)T_c$. In the case of quenched SU(3), \cite{NSS04} obtained using the data in the region of $T>1.5T_c$, $m_e/T=1.69(4)g(T)$ and $m_m/T=0.549(16)g^2(T)$.  Since the electric and the magnetic gluon propagator of \cite{NSS04} are normalized to 1 at $z=0$ and the critical temperature of the quenched configuration $T_c\sim 269\pm 1$MeV\cite{KaLa03} is much higher than that of the unquenched
configuration, we cannot compare quantitatively their data with ours, but their data of $T/T_c=1.32$ are consistent with ours within errors.  We do not observe suppression of $m_e/T$ near $T_c$. Whether the discrepancy is due to the presence of dynamical quarks is left for the future study.  
Discrepancy of about factor 6 in the $m_m/T$ of SU(3) near $T_c$ from the extrapolation of the pQCD results in $T>1.5 T_c$ region implies breakdown of the perturbation series near $T_c$.

\begin{figure}[htb]
\begin{center}
\includegraphics[width=7cm,angle=0,clip]{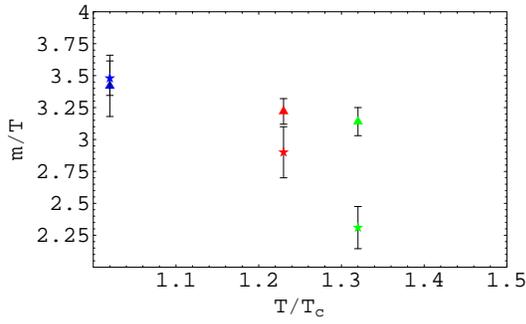}
\end{center}
\caption{The magnetic screening mass $m_m/T$ (stars) and the electric screening mass $m_e/T$ of MILC$_{ft}$ $N_f=2$ KS fermion unquenched configurations (triangles) . }\label{milcft_m}
\end{figure}

\section{The finite temperature QCD running coupling}\label{runningcoupling}

In \cite{GAMW04} the calculation of the running coupling in the Dyson-Schwinger equation(DSE) at zero temperature\cite{SHA} was extended to below $T_c$ and 
 the running coupling for $q_0=2\pi n T$ is written as 
\begin{equation}
\alpha(n, \Vec p^2,T)=\alpha(\mu^2)G^2(n,\Vec p^2,\mu^2,T)Z(n,\Vec p^2,\mu^2,T)
\end{equation}
where $G(n,\Vec p^2,\mu^2,T)$ is the ghost dressing function and $Z(n,\Vec p^2,\mu^2,T)$ is the gluon dressing function. Applying this method to $T\sim T_c$, we measure the QCD running coupling in the $\widetilde{MOM}$ scheme as the product of the gluon dressing function and the ghost dressing function squared, as a function of the momentum ${\Vec q}$ along the spatial axes.  We set $n$ equals 0 in the analysis.
We normalize the magnitude to the result of four-loop pQCD\cite{orsay01,ChRe00,chet,FN05b} at the highest momentum point of $\beta=5.725$.
A DSE result of zero-temperature quenched configuration with the infrared exponent of the ghost propagator $\kappa=0.5$ \cite{Bloch03}. In \cite{FN04} we compared the running coupling in the quenched lattice simulation with the parametrization suggested by the DSE analysis as
\begin{eqnarray}
&&\alpha_s(q^2)=\frac{1}{c_0+(q^2/\Lambda_{QCD}^2)^2}[c_0\alpha_0+\frac{4\pi}{\beta_0}\nonumber\\
&&\times \left(\frac{1}{\log(q^2/\Lambda_{QCD}^2)}-\frac{1}{(q^2/\Lambda_{QCD}^2)-1}\right)
(\frac{q^2}{\Lambda_{QCD}^2})^2]\nonumber\\
\end{eqnarray}
where $\beta_0=11-\frac{2}{3}N_f$.

\begin{figure}[htb]
\begin{center}
\includegraphics[width=7cm,angle=0,clip]{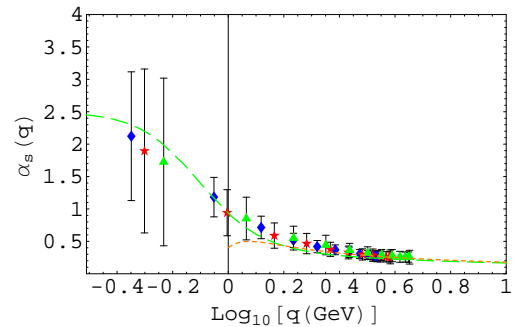}
\end{center}
\caption{The running coupling of the MILC$_{ft}$ of $\beta=5.65$(blue diamonds), 5.725(red stars) and 5.85(green triangles) as the function of $\log \Vec q$(GeV). Dashed line is the DSE result used to fit the quenched zero temperature lattice data \cite{FN04} with $\kappa=0.5$ and the dotted line is the pQCD result with $N_f=2$ of zero temperature. }\label{milcft_alpfit}
\end{figure}

The parameters adopted in the fit was $\Lambda_{QCD}=330$MeV, $c_0=30$. 
The $N_f$ dependence of the running coupling is not so large and to make comparison with the quenched data, the dashed line 
used to fit the quenched data of $N_f=0$ \cite{FN04} is shown in Fig.\ref{milcft_alpfit}. 

In the pQCD, an approximate inversion of the four-loop formula yields the running coupling as a function of $t=\log(q^2/\Lambda^2)$ as follows\cite{ChRe00,orsay01}.

\begin{eqnarray}
&&\alpha_{s,pert}(q^2)=\frac{4\pi}{\beta_0 t}-\frac{8\pi\tilde\beta_1}{\beta_0}\frac{\log (t)}{(\beta_0 t)^2}\nonumber\\
&&+\frac{1}{(\beta_0t)^3}\left(\frac{2\pi \tilde\beta_2}{\beta_0}+\frac{16\pi\tilde\tilde\beta_1^2}{\beta_0^2}(\log^2(t)-\log(t)-1)\right)\nonumber\\
&&+\frac{1}{(\beta_0t)^4}\left[\frac{2\pi\tilde\beta_3}{\beta_0}+\frac{16\pi\tilde\beta_1^3}{\beta_0^3}\left(-2\log^3(t)+5\log^2(t)\right.\right.\nonumber\\
&&\left.\left.+(4-\frac{3\tilde\beta_2\beta_0}{4\tilde\beta_1^2})\log(t)-1\right)\right]
\end{eqnarray}
where $\tilde\beta_1=(102-\frac{38}{3}N_f)/2$,
 and 

\noindent$\displaystyle \tilde\beta_2 = (\frac{2857}{2} - \frac{5033}{18} N_f + \frac{325}{54} N_f^2)/4$,

\noindent$\displaystyle \tilde\beta_3 = (\frac{149753}{6} + 
          3564 \zeta(3) + (-\frac{1078361}{162}  - 
              \frac{6508}{27} \zeta(3)) N_f$
\noindent$\displaystyle+(\frac{50065}{162}+\frac{6472}{81}\zeta(3)) N_f^2+
   \frac{1093}{729} N_f^3)/8$.

Here the scale in $\widetilde{MOM}$ scheme is defined as\cite{CeGo79,orsay98} 
\[
\Lambda=\Lambda_{\overline{MS}}e^{(70/3-22 N_f/9)/22}
\]
and $\Lambda_{\overline{MS}}$ is  0.259(22) GeV for $N_f=2$ and
0.252(10) GeV for $N_f=0$ were obtained by fitting the data in the continuum window 1.8GeV $<q<$2.3GeV ($0.26<\log_{10}[q({\rm GeV})]<0.37$)\cite{orsay01,orsay98}. 

The pQCD running coupling in the $\widetilde{MOM}$ scheme using $\Lambda_{\overline{MS}}=0.259$ GeV is shown by the dotted line.

Comparing with the result of $N_f=2+1$ zero-temperature\cite{FN05b}, the deviation from the pQCD is smaller.  Although the presence of ghost condensate at zero temperature is not excluded, there is no sign of finite $v$ and $A^2$ at finite temperature.

The running couplings of three temperatures near $T_c$ as a function of the momentum $\Vec q$ lies roughly on a single curve and there is not strong temperature dependence. The relatively large $T$ dependence of the $m_m/T$ suggests that the non-perturbative effects on the magnetic screening mass is important. 

The running coupling of zero temperature $N_f=2$ MILC $20^3\times 64$ configuration is roughly the same as that of $N_f=2+1$ MILC$_c$ presented in Fig.15 of \cite{FN05b}.  In the $-0.9<\log_{10}[q({\rm GeV})]<-0.3$ region it decreases monotonically.  We think that the suppression of the running coupling at the
infrared is due to the singularity of the color antisymmetric ghost dressing fuction there, which suppresses the color diagonal ghost dressing function, and that the suppression is a kind of an artefact. 
It may imply a question on the structure of the infrared fixed point and on the extensibility from the perturbative region to the non-perturbative region of the method of defining the running coupling from the product of the color diagonal ghost dressing function squared and the gluon dressing function. 

\section{Discussion and conclusion}
We measured the color diagonal and the color antisymmetric ghost propagator of quenched, unquenched zero temperature and unquenched finite temperature configurations. 
The quark has the effect of quenching randomness of the ghost propagator and enhancing the transverse and magnetic gluon propagator above $T_c$.
 The Binder cumulant of the quenched SU(3) color antisymmetric ghost propagator is noisy, but that of the unquenched SU(3) color antisymmetric ghost propagator is almost constant and independent of the momentum except the lowest momentum point.
 The color diagonal ghost propagator of finite temperatures above $T_c$ are almost temperature independent, but the scale of the color antisymmetric ghost propagator becomes larger as the temperature becomes higher. 

We observed that the Binder cumulants of zero temperature unquenched $N_f=2+1$ SU($N_c$) configurations are consistent with those of $N_c^2-1$ dimensional Gaussian distributions. 
 The Binder cumulants of finite temperature unquenched $N_f=2+1$ SU(3) configurations deviate from Gaussian distribution as the temperature rises, which may be interpreted as the quark effect of quenching randomness is reduced in high temperature.
 The anomaly of the KS fermion propagator and the anomaly of the momentum dependence of the Binder cumulant seems to be correlated. It may indicate that the QCD-like region of $N_f=2+1$ KS fermion system is more complicated than that of the $N_f=2$ case given in \cite{HH06}. The Binder cumulant of unquenched $N_f=2$ and 3 which are consistent with 0.66 indicates that the ratio of the $ud-$quark mass and the $s-$quark mass is an important ingredient. 

The parameter $v$ introduced to investigate the ghost condensate is found to be small and its value depends on the temperature and the bare mass of the KS fermion. 

We observed strong non-perturbative effects in the magnetic screening mass of the gluon near $T_c$. Since near $T_c$, systematic perturbative QCD calculation is impossible, it is important to formulate the lattice theory. In the quenched finite temperature Landau gauge DSE approach \cite{GAMW04,MWA05} the infrared exponent $\kappa$ and the running coupling in the MOM scheme were found to be essentially temperature independent below $T_c$.

\begin{acknowledgments}
We thank F. Karsch for a discussion on the scale assignment of the lattice data and A. Nakamura and T. Saito for the information on the quenched simulation of their group.
This work is supported by the High Energy Accelerator Research Organization(KEK) supercomputer project No. 05-128 using Hitachi-SR8000 and the large scale simulation program No.5(FY2006) using SR11000.  Numerical calculation of the ghost propagator was carried out by NEC-SX5 at the CMC of Osaka University and by NEC-SX8 at the Yukawa Institute Computer Facility in Kyoto University. H.N. is supported by the MEXT grant in aid of scientific research in priority area No.13135210. 
\end{acknowledgments}

\end{document}